# Driver perceptions of advanced driver assistance systems and safety

Sophie Le Page, Jason Millar, Kelly Bronson, Shalaleh Rismani, and AJung Moon

*Abstract*—Advanced driver assistance systems (ADAS) are often used in the automotive industry to highlight innovative improvements in vehicle safety. However, today it is unclear whether certain automation (e.g., adaptive cruise control, lane keeping, parking assist) increases safety of our roads. In this paper, we investigate driver awareness, use, perceived safety, knowledge, training, and attitudes toward ADAS with different automation systems/features. Results of our online survey (n=1018) reveal that there is a significant difference in frequency of use and perceived safety for different ADAS features. Furthermore, we find that at least 70% of drivers activate an ADAS feature "most or all of the time" when driving, yet we find that at least 40% of drivers report feeling that ADAS often compromises their safety when activated. We also find that most respondents learn how to use ADAS in their vehicles by trying it out on the road by themselves, rather than through any formal driver education and training. These results may mirror how certain ADAS features are often activated by default resulting in high usage rates. These results also suggest a lack of driver training and education for safely interacting with, and operating, ADAS, such as turning off systems/features. These findings contribute to a critical discussion about the overall safety implications of current ADAS, especially as they enable higher-level automation features to creep into personal vehicles without a lockstep response in training, regulation, and policy.

## I. INTRODUCTION

Advanced driver assistance systems (ADAS) are designed to improve vehicle safety, yet there are many factors that can affect the overall safety impact of the system. With the introduction of any novel vehicle technology, especially those that perform safety-critical driving tasks, a lack of driver trust in, and effective use of, the technology can make the overall safety impact on our roads negligible. Over reliance on, or ineffective use of such systems, on the other hand, can lead to worse safety outcomes. As many ADAS systems are being introduced into vehicles, some researchers have begun investigating the impact of automation on road safety by testing and evaluating the systems within simulated traffic scenarios such as lane changes [12], cut-ins [8], and collision avoidance [11]. Only recently is naturalistic crash data being analyzed to better understand the safety impact of varying degrees of automation in personal vehicles. Goodall [7], for example, found that autonomous vehicles are involved in more struck from behind crashes per vehicle-mile travelled compared to human-driven vehicles. His surprising results suggest we should not assume that automated driving systems improve safety. By extension, without data we should be skeptical as to whether current ADAS improves road safety.

In this study we use driver testimony as a starting point to understand the impact of SAE Level 1/2 automation systems on road safety.[1] We address the following research questions: (1) To what extent are drivers aware of the various ADAS in their vehicles?; (2) How frequently do drivers use ADAS in their vehicles?; (3) How often do drivers feel their safety is compromised when they use ADAS?; (4) How knowledgeable are drivers about ADAS in their vehicles?; (5) What type of ADAS training do drivers receive?; and (6) What are drivers' positive and negative attitudes towards ADAS and towards government testing of drivers? We conducted an online survey asking participants to reflect on ADAS in SAE Level 1/2 (e.g., adaptive cruise control, lane keeping, park assist) and SAE Level 0 (e.g., cruise control), where Level 0 (lower-level) serves as our "control" feature to compare to the more advanced Level 1/2 (higher-level) features.

## II. BACKGROUND

### A. Crash data and safety of ADAS

What do data on vehicle crash events reveal about ADAS safety? While Goodall's work [7] focused only on the safety implications of fully autonomous vehicles, other studies investigated safety implications of lower-level ADAS using crash data. For instance, studies by Cicchino suggest that lane departure warnings (LDW) reduced police-reported crashes by 18% compared to vehicles without LDW [4]. In another study, Cicchino found that the combination of rear emergency parking braking, rear vision cameras and rear parking sensors were found to reduce police-reported backing crashes by 78% compared to vehicles without those systems [5].

Kim et al. [9], on the other hand, analyzed safety critical events (SCEs) showing that higher-level ADAS safety could be improved both technically (e.g., ADAS disengaging when encountering blurred lane markings), and from a human factors perspective (e.g., monitoring drivers to reduce distraction/engaging in secondary tasks). They find that 20% of SCEs involved ADAS being turned on, and 57% of those involved people ineffectively using the ADAS (e.g., engaged in secondary tasks, using the systems not on highways, or

S. Le Page is with the School Engineering Design and Teaching Innovation, University of Ottawa, Ottawa, Canada (e-mail: slepa017@uottawa.ca).
J. Millar is with the School Engineering Design and Teaching Innovation, University of Ottawa, Ottawa, Canada (e-mail: jmillar@uottawa.ca).
K. Bronson is with the Faculty of Social Science, University of Ottawa, Ottawa, Canada (e-mail: kbronson@uottawa.ca).
S. Rismani is with the Open Roboethics Institute (e-mail: shalaleh@openroboethics.org).
A. Moon is with is with the Faculty of Engineering, McGill University, Montreal, Canada (e-mail: ajung.moon@mcgill.ca).

---

[1] SAE has proposed an influential categorization for "levels of driving automation" in vehicular automation.

with hands off the wheel). During 13% of SCEs, the ADASs neither reacted to the situation nor warned the driver. Their work highlights human factors as one of the main challenges with safe use of ADAS.

Crash data that provides a detailed understanding of the safety implications of higher-level ADAS is hard to come by. We found very little research on naturalistic driving data (e.g., crash rates) showing that sustained longitudinal (i.e., acceleration) and lateral (i.e., steering) control systems (e.g., adaptive cruise control, lane keeping assist) are safer than human-driven vehicles. We found one study shared with us by Winkle et al. [17] who performed Audi pre-crash data analysis, finding adaptive cruise control and lane departure warning helped to avoid accidents and reduce severity. Given the lack of existing ADAS safety data, an alternative methodology for investigating higher-level ADAS's impact on safety is to ask drivers about the nature of their experiences using ADAS.

*B. SAE Automation Levels and ADAS*

SAE J3016 defines six levels of automation for cars, ranging from level 0 (No Driving Automation) to level 5 (Full Automation) [18]. This categorization scheme has also been adopted by the NHTSA. To be classified, at a minimum a system must include sustained lateral or longitudinal control. Since anti-lock brake systems, electronic stability control, or automated emergency braking do not perform such control on a sustained basis, they are not classifiable (other than level 0). Conventional cruise control does not provide sustained operation because it does not respond to external events. It is therefore also not classifiable (other than at level 0). Adaptive cruise control (ACC) which performs longitudinal control and lane keeping assist (LKA) which performs lateral control, are classified as Level 1. Level 2 systems perform both lateral and longitudinal control e.g., systems using both ACC and LKA at once, or parking assist systems that will parallel park your vehicle hands-free. If a driver does not need to monitor a system and will be alarmed when to intervene, the system is considered Level 3. If a driver never needs to intervene in certain areas, the system is Level 4, and in all areas the system is Level 5.

*C. Driver perceptions and safety of ADAS*

National surveys related to drivers' perceptions of SAE Level 1/2 ADAS have been conducted before. Eichelberger and McCartt conducted the first US study on the use of adaptive cruise control (ACC) and lane departure warning/prevention in a non-luxury vehicle (Toyota) [6]. They found that consumer acceptance of these technologies was generally high (90% wanted ACC again), although less so for lane departure warning/prevention (71%). Eichelberger and McCartt also find that for Toyota drivers in the US there have been concerns that technologies may have unintended effects on driver behavior, or might make drivers less attentive. In their study some drivers reported that they allowed the vehicle to brake for them in situations where the system determined that emergency braking was needed to avoid a crash. Eleven percent of drivers who used ACC reported that they usually waited for the warning before braking as they approached another vehicle. Four percent of those with ACC said that they looked away from the road more often when using ACC. For lane keeping assist 9% mentioned that it was distracting (1% for ACC). Their findings suggest that drivers knowingly adopt potentially unsafe driving practices when ADAS are activated, which raises safety concerns.

A number of national surveys across the world suggest that driver awareness and training about ADAS are related to driver acceptance of the systems. For instance, Abraham et al. [1] look at how drivers learn, and would prefer to learn, about ADAS in their vehicles in the US, finding that drivers who learned through their preferred method – dealership training was the most preferred (55%) – reported higher understanding and use of the systems. Boelhouwer et al. found that in the Netherlands almost 40% of the car dealers did not receive (sufficient) information about ADAS and almost a quarter of drivers did not receive information about ADAS in the car they bought [2]. Of the drivers who did receive information only 9% were able to try the system before taking the car home. Oxely et al. [13] look at whether Australian drivers know of, have, and are willing to pay for features such as ACC and lane keeping assist, finding an overall low awareness of the features (28.5% responded knowing what ACC is, 43.1% for lane change warning/assist). Of those who knew what the feature was, the authors also found an overall low availability of the features (44.1% responded that their vehicle is fitted with ACC, 19.9% for lane change warning/assist). On the other hand, the authors found there was greater reported interest in purchasing the features (64.3% responded ACC would be a priority when purchasing their next vehicle, 60.7% for lane change warning/assist).

Surveys across the world also suggest respondents feel unsafe using ADAS in vehicles. Lijarcio et al. [10] look at how often Spanish drivers use ADAS features, as well as reasons drivers do not use these features. Options include "I don't need it/find it useless", "I don't know how it works", "I don't trust this feature", "It distracts me", and "I find it annoying/uncomfortable". Respondents reported not using a feature due to not trusting it highest for "automatic parking systems" (34.2% due to "I do not trust it", 26.3% due to "it distracts me"), followed by lane keeping assistant (18.8% "not trusting", 33.3% "it distracts"), and ACC (15.2% "not trusting", 16.2% "it distracts"). Overall, the authors found that low perceived value, lack of confidence and potential distractibility are the main constraints perceived by drivers to use ADAS features. In Germany, Brell et al. [3] compare German drivers' risk perception of vehicles with and without ACC and found that there is a higher perceived risk of vehicles with ACC than without, and that this perceived risk reduces with actual use of ACC compared to participants with no experience.

As far as we are aware, ours is the first US national survey on drivers' perceived safety when using ADAS that take over longitudinal or lateral control (e.g., ACC, LKA, park assist). Furthermore, this work contributes to understanding how driver use, knowledge, training and attitudes correlate with their ADAS safety perceptions.

## III. METHODS

### A. Participants

A total of 1018 vehicle-owners from the United States (gender and age census balanced, 536 females, 482 males) completed an online survey in June 2020. SurveyMonkey Target Audience service was used to recruit target participants where the participants took surveys for charity and a chance to win sweepstake prizes distributed by SurveyMonkey.

Most respondents have over 10 years of experience driving (68.4%), followed by 5-10 years (15.6%), and less than 5 years (13.6%). Some participants (2.4%) reported "having never driven" despite reported as owning a vehicle. This may indicate that while the participants own a vehicle, they are primarily vehicle passengers as was found to be the case for 8.1% of participants in Brell et al. [3].

### B. Materials and procedure

The survey consisted of 22 multiple choice questions, where some multiple-choice questions were randomized to account for possible order effects. It took around 10 minutes for participants to complete the survey. The survey is available at https://openroboethics.org/ori-koti-2020-public-opinion-poll-data/.

After providing information about their duration of driving experience, participants were asked whether they have, frequently use (*Frequency of Use*), or feel their safety is compromised (*Perceived Safety*) when using an ADAS. Participants provided answers in a 5-point Likert scale, where for *Frequency of Use*, 1 indicated "never", 2 indicated "rarely", 3 indicated "most of the time when I drive", 4 indicated "every time I drive", and 0 was reserved for "I don't know" responses. The 5-point Likert scale was similar for *Perceived Safety* except for 3 indicating "most of the time when I use the feature", and 4 indicating "every time I use the feature". The same questions were asked across the following *ADAS types*: cruise control, adaptive cruise control (ACC), lane keeping (LKA), and parking assist. Subsequently, participants were asked about their knowledge of, and methods for, learning about ADAS, with additional questions about the duration and quality of education or training received from the dealership. Lastly, the participants were asked about their general positive and negative attitudes towards ADAS and government testing of drivers.

We highlight only the key results from the survey in this paper. Full survey results including participant's attitudes towards ADAS can be found on our website (provided above).

### C. Analysis

We conducted a two-way multivariate analysis (MANOVA) to investigate whether there were any significant differences in *Frequency of Use* and *Perceived Safety* as rated by the participants across factors *Gender* and *ADAS type*. For the MANOVA, "I don't know" responses were removed. We first investigated whether there were any significant interaction effects between gender and ADAS type on the two dependent variables (*Frequency of Use* and *Perceived Safety*). We followed with a number of analyses of variance (ANOVA) to study the main effects from *Gender* and *ADAS type* on *Frequency of Use* and *Perceived Safety*. Table I, II and III contain the basic descriptive statistics on our data (where "I don't know" responses were removed).

TABLE I. Number of data per independent variable

| | Value label | N |
|---|---|---|
| ADAS Type | Cruise Control | 862 |
| | Adaptive cruise control | 279 |
| | Lane keeping | 246 |
| | Parking assist | 196 |
| Gender | Male | 787 |
| | Female | 796 |

TABLE I. Descriptive statistics for *Frequency of Use*

| ADAS Type | Gender | Mean | Std. Dev. | N |
|---|---|---|---|---|
| **Cruise control** | Male | 2.44 | 0.734 | 418 |
| | Female | 2.27 | 0.740 | 444 |
| **Adaptive cruise control** | Male | 2.56 | 0.736 | 151 |
| | Female | 2.41 | 0.892 | 128 |
| **Lane keeping** | Male | 2.92 | 0.925 | 124 |
| | Female | 3.18 | 0.843 | 122 |
| **Parking assist** | Male | 2.85 | 0.915 | 94 |
| | Female | 3.00 | 0.901 | 102 |
| **Total** | Male | 2.59 | 0.813 | 787 |
| | Female | 2.53 | 0.880 | 796 |

TABLE II. Descriptive statistics for *Perceived Safety*

| ADAS Type | Gender | Mean | Std. Dev. | N |
|---|---|---|---|---|
| **Cruise control** | Male | 1.83 | 0.778 | 418 |
| | Female | 1.82 | 0.829 | 444 |
| **Adaptive cruise control** | Male | 2.09 | 0.919 | 151 |
| | Female | 2.05 | 0.987 | 128 |
| **Lane keeping** | Male | 2.19 | 0.985 | 124 |
| | Female | 2.16 | 1.063 | 122 |
| **Parking assist** | Male | 2.33 | 1.061 | 94 |
| | Female | 2.16 | 1.097 | 101 |

| Total | Male | 2.00 | 0.896 | 787 |
|---|---|---|---|---|
| | Female | 1.95 | 0.942 | 796 |

Our assumption testing for the MANOVA revealed that the independent and dependent variables did not meet univariate normality assumption (Shapiro-Wilks, $p > 0.05$). However, since each independent variable has more than 20 data points, the MANOVA results are still robust to this violation [15]. The assumption of homogeneity of variance and variance-covariance are also violated based on the results of the Box's and Levene's tests. To account for this violation, we report the Pillai's Trace results. All other MANOVA assumptions were tenable.

## IV. RESULTS

### A. Awareness of ADAS

Respondents were asked if they know whether the vehicle they most often drive has various ADAS features. A majority of participants indicated having cruise control (89.3%, $n=919$), followed by 28.5% having ACC ($n=293$), 25.0% having LKA ($n=257$), and 20.3% having parking assist ($n=209$). Respondents were most unsure as to whether their car has ACC (35.8% responding "I don't know"), followed by LKA (8.9%), parking assist (4.6%), and cruise control (3.1%).

### B. Frequency of use and perceived safety of ADAS

Of the respondents who reported having ADAS features in their vehicles, the respondents were further asked to indicate how often they turn on (i.e., activate) the individual features. The majority (71.2%) of respondents turned on lane keeping "most or all of the time" or "every time I drive", followed by parking assist (63.2%), adaptive cruise control (45.1%), and cruise control (36.6%). The number of respondents to "never" activate a feature was approximately the same for all of the features (12.0% for cruise control, 11.3% for ACC, 9.6% for parking assist), except for lane keeping, where half as many respondents "never" activate lane keeping compared to the other features (5.5% for LKA).

Respondents who reported having the ADAS feature in their vehicles were also asked whether they had ever felt their safety was compromised when using the feature. 40.7% of respondents felt that parking assist compromised their safety "most or all of the time" when activated, followed by lane keeping (37.0%), ACC (30.4%), and cruise control (16.5%). The "rarely" options are consistent in the opposite direction, where more respondents (41.1%) felt that cruise control "rarely" compromised their safety compared to parking assist (21.1%). Furthermore, twice as many respondents felt adaptive cruise control compromised safety "most or all of the time" when activated compared to cruise control (30.4% vs 16.5%). The number of respondents to report a feature "never" compromised their safety when activated was approximately the same for all of the features.

Comparing "frequency of feature use" and "frequency of feeling that feature use compromises safety", generally the more an ADAS feature was activated "most or all of the time", the more the drivers experienced feeling the feature, when activated, compromised safety "most or all of the time". An exception is parking assist: even though parking assist was turned on less often than lane keeping (63.2% vs 71.2% responded turning on the feature "most or all of the time") it was felt to compromise safety more "most or all of the time" when activated (40.7% vs 37.0%).

Results of the MANOVA indicate that there is a significant interaction effect of *Gender* and *ADAS type* on *Frequency of Use* and *Perceived Safety* (*Pillai's Trace = 0.12, F(6, 3150) =3.27, p = 0.003*). The effect size is small and this is likely due to different number of data points across independent variables. The power is 0.935 which means that the results are significant 93.5% of the time. Based on these results, we can reject the null hypothesis.

Figure 1. Frequency of use for each ADAS type across the two genders.

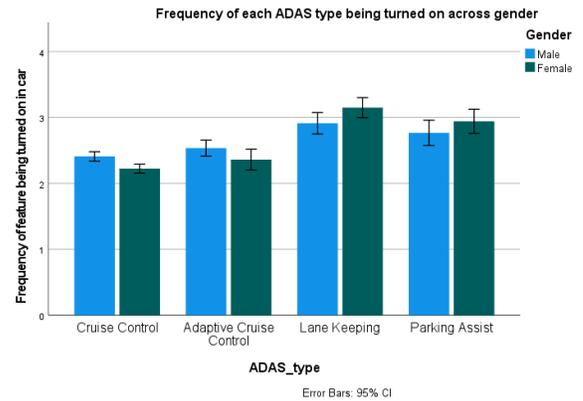

Figure 2. Perceived safety for each ADAS type across the two genders.

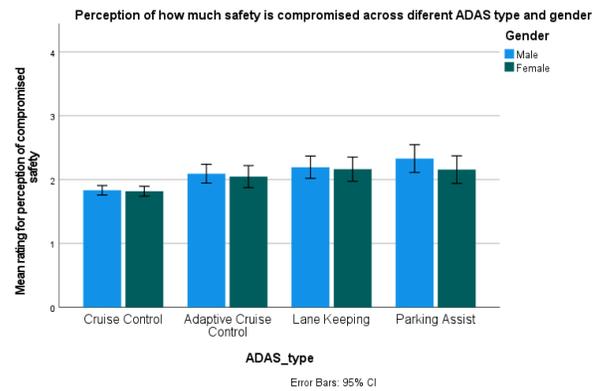

After the initial MANOVA test, we conducted a number of ANOVA to understand the relationship between the variables considering the significant interaction effect. Using the Bonferroni method each ANOVA was tested at $a = 0.025$. The results of ANOVA indicate that the interaction effect of *Gender* and *ADAS type* are significant for *Frequency of Use*, *F(3, 1575) = 5.89, p < 0.001*. The interaction effect is not significant for *Perceived Safety*, *F(3, 1575) = 0.41, p = 0.746*. The ANOVA also shows that there are significant main effects of *ADAS type* on *Frequency of Use*, *F(3, 1575) = 64.23, p < 0.001* and on *Perceived Safety*, *F(3, 1575) = 18.79, p < 0.001*. However, we see no main effects from *Gender* on *Frequency of Use* and *Perceived Safety*. This means that one gender uses a type of ADAS feature more frequently than another, even though their perception of safety for each ADAS feature isn't significantly different. Our analysis shows that male drivers (*M=2.4, SD=0.75*) use cruise control significantly more than female drivers (*M=2.2, SD=0.77*) (*p<.001*). We found a statistical pattern indicating female drivers (*M=3.4, SD=0.79*) might use lane keeping more than male drivers (*M=2.9, SD=0.96*), however, this relationship is not significant (*p=0.09*). There is a possible indication that male drivers (*M=2.29, SD=1.15*) might perceive parking assist to compromise their safety more than female drivers' perception (*M=1.64, SD=1.05*). However, this pattern is not statistically significant (*p=0.08*).

Figure 1 and Figure 2 illustrates some patterns that are indicative of the main and interaction effects we identified in our MANOVA and ANOVA.

We also investigated the connection between *Perceived Safety* and perceived knowledge of *ADAS type*. There is no clear correlation between these two variables. However, females consistently rank their perceived knowledge lower than that of male participants.

### C. Knowledge of, and training on, ADAS

Respondents were asked to indicate how they would rate their knowledge of the ADAS built into their car among four options: "little to no knowledge", "some knowledge", "considerable knowledge", or "expert knowledge". 35.0% of respondents rated their knowledge as "considerable" or "expert", 34.9% had "some knowledge", and 30.1% had "little to no knowledge".

Respondents were also asked to indicate what kinds of training or learning had contributed to their knowledge of ADAS in their vehicles among ten options: "trying the feature(s) out" (37.9%), "reading the car manual" (37.4%), "at the dealership" (27.8%), "through a website" (18.5%), "training videos in the vehicle" (8.5%), "a driving course" (7.7%), "a phone app" (9.1%), "at the department of motor vehicles" (5.8%), "from customer service" (4.0%), and "not applicable" (26.7%). Most respondents (37.9%) learned about these systems by trying them while driving. Only a small proportion (7.7%) took a driving course to learn about ADAS in their cars.

341 out of 1,020 (33.4%) respondents answered "yes" to receiving training from dealership/salespersons(s) about ADAS in their vehicles (39.5% responded "no", 27.1% "not applicable"), and about half of those respondents (152 out of 1,020) reported receiving "less than 10 minutes" of training (43.7% responded receiving "between 10 and 30 minutes", 6.8% "more than 30 minutes", 4.7% "none"). 77.6% of respondents rated the training as "very good" or "good", 20.9% rated the training as "very poor" or "poor", and 1.5% responded "I don't know".

## V. DISCUSSION

### A. Awareness of ADAS

Our findings suggest that about 1 in 4 drivers have a higher-level ADAS feature (e.g., adaptive cruise control, lane keeping, parking assist) in their vehicle, and that about 90% of drivers have lower-level ADAS features such as cruise control. These are consistent with findings from previous studies [2], [10], [13]. We also find that drivers are less aware of higher-level ADAS features such as ACC in their vehicles than lower-level features such as conventional cruise control. This is expected since cruise control has been a feature in vehicles for longer.

Our respondents were least sure about whether their vehicle had ACC. Respondent's high level of uncertainty may be due to a confusion between ACC and cruise control. For example, Tsapi [16] look at introducing ADAS into drivers' training and testing in the Netherlands and find that the ACC system was confused by more than half of the learner drivers with cruise control, creating a wrong impression of its capabilities and areas of application. Respondent's high uncertainty about having ACC may also be because ACC is named differently depending on the car brand. For example, the 2010–2013 Toyota Sienna and Prius models are equipped with "Dynamic Radar Cruise Control", another name for adaptive cruise control [6].

Note that we did not take the respondent's car brands and models into account in our survey and did not cross-check whether the participants are unaware of the ADAS that exist in their car. For further work to investigate and improve safety of ADAS, one could look at details of the manufacturer systems separately to account for different systems having varied impacts on respondents' understanding and experiences of the systems.

### B. Frequency of use and perceived safety of ADAS

We found that at least 70% of drivers activate an ADAS feature "most or all of the time" when driving, while at least 10% of drivers never activate an ADAS feature. We also found at least 40% our respondents reported feeling that ADAS often compromise their safety when activated. We also found that higher-level features such as ACC are perceived as less safe than lower-level features, such as cruise control. These findings are consistent with previous studies [6], [2], [10], [3].

That we found lane keeping was used more compared to other features could be due to default settings. For example, Eichelberger and McCartt [6] state that "differences in the reported use of […] systems could be due to the default settings" when they found that the reported usage of Toyota's lane-keeping assist systems that are turned off by default was lower than the usage of Volvo and Infiniti lane departure warning systems that are turned on by default.

Comparing our findings of feature usage rates to other studies, we found some differences: notably lane keeping and parking assist were reported as being used less by respondents in previous studies [6], [2], [10]. Aside from the variance in car manufacturer feature interfaces and default settings accounting for the differences in usage rates across studies, the differences may also be due to how the features are described in the surveys. When we conducted our survey, to avoid influencing respondent's understandings and awareness of these features in their vehicles, we did not provide descriptions of the features. As a result, respondents may have understood lane keeping assist and parking assist to include lower-level warnings and sensor features that do not take over driving control, resulting in higher reported usage rates from our respondents compared to previous studies. In contrast, Boelhouwer et al. (2020) described lane keeping as

"This system keeps the car within the lane. The car steers itself to keep between the lane markings." and described automated parking as "The car parks itself". Of course, these differences in manufacturer ADAS features, in other words the lack of standardization across ADAS features, could further contribute to driver confusion regarding ADAS functionality.

Comparing our findings of perceived safety of ADAS to previous studies, we found similarities in drivers reporting feeling unsafe when using the systems: Lijarcio et al. [10] found 34.2% of respondents did not use automated parking due to not trusting it, followed by LKA (18%) and ACC (15.2%), while Brell et al. [3] found a higher perceived risk of vehicles with ACC than without (such as vehicles with conventional cruise control). However, Lijarcio et al. do not report on whether drivers have experienced using the feature, while Brell et al. do not report on how much drivers experience using a feature. Our survey asks drivers to directly report their experience and on how much they have used the feature, which we suspect is a more accurate read on the perceived safety of a system.

The fact that a large percentage of licensed drivers are reporting feeling that ADAS compromise their safety "most or all of the time" when activated is alarming. It is evidence that strongly suggests ADAS does not straightforwardly improve safety. At the very least it suggests the urgent need for detailed naturalistic driving data studies into the impacts of ADAS on safety, especially since the automotive industry is on a trajectory to introduce more, and increasingly complex, ADAS to market in the coming years.

Future work and methodological rigor would be needed to understand use and driver perceptions of safety of ADAS. We believe we have conducted the first survey of its kind. However, because of the size of our survey we did not design the survey to focus solely on perceived safety of ADAS. Future work could address this.

*C. Knowledge of, and training on, ADAS*

We found that 30.1% of respondents had little to no knowledge of ADAS. Only 33.4% of respondents received instruction from the dealership about these systems and most learned by using these systems while driving. Furthermore only 7.7% of respondents reported taking a driving course to learn about ADAS in their cars.

Our findings are similar to Boelhouwer et al. [2], who found almost a quarter of respondents in the Netherlands did not receive information on any systems at the car dealer. Abraham et al. [1] also conducted a large US survey of 2,364 participants to determine drivers' used and preferred methods for understanding ADAS in their vehicles, finding 12% of participants received no explanation at the dealership when purchasing a vehicle with ADAS. Similar to our findings, both Boelhouwer et al. [2] and Abraham et al. [1] also find that most drivers got the information either through the owner's manual, or by trial-and-error practice while driving. Abraham et al. [1] further find that the preferred way of learning about ADAS is through the dealership.

However, dealership "training" may give drivers a false sense of security since salespersons are not actually "training" drivers. That is, dealerships are not preparing drivers for licensing or testing on ADAS, nor are they testing drivers' ability to effectively operate ADAS, calling into question whether this qualifies as "training" per se. If drivers are feeling unsafe, as our findings suggest, it could be from a clear lack of driver training with corresponding licensing and testing requirements. ADAS are becoming more prevalent in vehicles, and those systems require a novel driver activity that is more akin to supervising/monitoring the driving task for their safe and effective use. It stands to reason that driver licensing should include testing a would-be driver's ability to operate these increasingly complex systems safely and effectively.

VI. CONCLUSION

ADAS increasingly perform safety-critical driving tasks, so it is reasonable to assert that drivers should be properly trained to operate these systems. Our findings suggest that drivers are not so trained. To our knowledge no formal training or licensing requirements on existing or emerging ADAS technologies currently exist [14]. Therefore, driver training should be supplemented with new material to ensure the safe and effective operation of ADAS on specific car manufacturer systems.

A large percentage of our respondents reported feeling that ADAS often compromise their safety. Thus, we conclude that more research should be done, including naturalistic driving data and human factors, to better understand the full safety impact of ADAS. Regulations should also be imposed on manufacturers to provide transparent information about the use and operational limitations of ADAS [14]. The implications of our findings also point to the need for more research to understand driver perceptions of safety for autonomous vehicles (than just ADAS) to correspond with the crashes and safety data (cited above).

Furthermore, our finding that higher-level features are perceived as less safe than lower-level features, such as cruise control, suggests either a problem with ADAS system design, or with driver training, or both. It also suggests that the current automotive trajectory—to increase the complexity of ADAS—is unsustainable. In short, if the going assumption is that putting ADAS in cars makes mobility safer, a responsible approach to ADAS development requires us to reexamine that assumption.

ACKNOWLEDGMENT

This research was approved by the Health Sciences and Science Research Ethics Board of the University of Ottawa (REB approval number H-05-20-5780). This research was generously supported by a grant (code 20TLRP-B127858-04) from the Korean Ministry of Land, Infrastructure and Transport's Transportation and Logistics R&D Program. It was also generously supported by the Open Roboethics Institute, and with funding through the Social Sciences and Humanities Research Council of Canada's (SSHRC) Canada Research Chairs program.